\documentclass[aps,prl,twocolumn,superscriptaddress,showpacs,floatfix]{revtex4-1}
\usepackage{graphicx,epsfig,amsmath}

\newcommand{\etal}{\textit{et al.}}

\newcommand{\esqoh}{\mbox{$e^2/h$}}

\newcommand{\Startsubfig}[2]{Figure~\ref{fig:#1}(#2)}
\newcommand{\subfig}[2]{Fig.~\ref{fig:#1}(#2)}
\newcommand{\allfig}[1]{Fig.~\ref{fig:#1}}

\newcommand{\didv}{\mbox{$\text{dI}/\text{dV}_{\text{ds}}$}}

\newcommand{\AsubB}[2]{\mbox{$\text{#1}_{\text{#2}}$}}

\begin{document}


\title{Coulomb Blockade in an Open Quantum Dot}

\author{S. Amasha}
	\email{samasha@stanford.edu}
	\affiliation{Department of Physics, Stanford University, Stanford, California 94305, USA}

\author{I. G. Rau}
	\affiliation{Department of Applied Physics, Stanford University, Stanford, California 94305, USA}

\author{M. Grobis}
	\altaffiliation{Present address: Hitachi GST, San Jose, CA 95135}
	\affiliation{Department of Physics, Stanford University, Stanford, California 94305, USA}

\author{R. M. Potok}
	\altaffiliation{Present address: Solyndra, Fremont, CA 94538}
	\affiliation{Department of Physics, Stanford University, Stanford, California 94305, USA}
	\affiliation{Department of Physics, Harvard University, Cambridge, Massachusetts 02138, USA}

\author{H. Shtrikman} 
	\affiliation{Department of Condensed Matter Physics, Weizmann Institute of Science, Rehovot 96100, Israel}
	
\author{D. Goldhaber-Gordon} 
	\affiliation{Department of Physics, Stanford University, Stanford, California 94305, USA}	


\begin{abstract}

   We report the observation of Coulomb blockade in a quantum dot contacted by two quantum point contacts each with a single fully-transmitting mode, a system previously thought to be well described without invoking Coulomb interactions. At temperatures below $50~\mbox{mK}$ we observe a periodic oscillation in the conductance of the dot with gate voltage that corresponds to a residual quantization of charge. From the temperature and magnetic field dependence, we infer the oscillations are Mesoscopic Coulomb Blockade, a type of Coulomb blockade caused by electron interference in an otherwise open system. 

\end{abstract}


\maketitle


   Mesoscopic systems are conventionally divided into two classes. In closed systems electrons are localized and Coulomb interaction effects determine the transport properties, while in open systems the Coulomb interaction can be neglected at low energies. The class of a system is thought to depend on the contacts between the mesoscopic region and the surrounding electrons. If the contacts contain a large number of poorly transmitting channels, such as in metallic nanostructures, then the crossover from closed to open is smooth and occurs when the total conductance of the contacts is on the order of $\esqoh$ \cite{Joyez1997:StrongTunneling, Grabert1992:SingleChgTunneling}. If the contacts each have one mode, such as can happen in semiconductor nanostructures, then the transition from closed to open is sharp: Coulomb blockade occurs when the mode in each contact is partially transmitting, and in the absence of phase coherence Coulomb blockade disappears when the mode in either contact becomes fully transmitting \cite{Matveev1995:CBatPerfectT}.
   
   This transition from the closed to the open regime has been demonstrated with laterally gated quantum dots \cite{Kouwenhoven1991:SingleChgEffects, vanderVaart1993:ChgInHighB, Pasquier1993:QLimitOnCB}. Such a dot is contacted via one-dimensional channels called quantum point contacts (QPCs). A QPC is tunable, and its conductance $G_{QPC}$ is directly related to the transmission of its modes: $G_{QPC}= 2 \esqoh$ corresponds to a single fully transmitting spin-degenerate mode, while  $G_{QPC} \ll 2~\esqoh$ corresponds to the tunneling regime. A dot is typically contacted by two QPCs. As the conductance of either QPC increases, the energy to add an additional electron to the dot (the charging energy $U$) is reduced \cite{Flensberg1994:DotsConToQPCs, Molenkamp1995:CoulScaling} and capacitance measurements show that the Coulomb oscillations decrease in amplitude, disappearing entirely at $G_{QPC}= 2~\esqoh$ \cite{Berman1999:QuantFlucOfChg, Duncan1999:DestroyQquant}.   

    Phase coherence complicates the transition from closed to open. In a one-leaded dot (a dot where the conductance of one QPC is adjustable, while the other is kept $\ll 2~\esqoh$) Coulomb blockade features are observed even if the adjustable QPC is set at $2~\esqoh$ \cite{Cronenwett1998:MCBin1ChDots}. Coherent electron paths in the dot interfere at this QPC and can reduce its transmission, trapping electrons on the dot. This leads to a type of Coulomb blockade called Mesoscopic Coulomb Blockade (MCB) \cite{Aleiner1998:MCB}. Electron interference, and hence MCB, is strongest at zero magnetic field because a closed path that begins and ends at the same QPC interferes constructively with its time-reversed pair, an effect called Weak Localization (WL).  In contrast to the one-leaded case, a coherent dot where both QPCs have many fully transmitting modes is predicted not to have MCB \cite{Brouwer2005:NonequilCBThy, Golubev2004:ElecTransportThruQD}: there are now multiple escape paths and it is unlikely that interference will reduce the transmission of all paths simultaneously. This result is expected to extend to the case where each QPC has just one fully transmitting spin-degenerate mode and MCB should be absent. To date transport measurements have confirmed that this system is open: while there are hints of MCB \cite{Pasquier1993:QLimitOnCB, Huibers1999:Thesis}, most experimental results \cite{Chan1995:CondFluc, Huibers1998:CondDistributions} have been well understood using Random Matrix theory that neglects explicit Coulomb interactions \cite{RMTpapers}.

   In this Letter, we report measurements of MCB in a dot where each QPC has a fully transmitting mode. We reach an electron temperature of $13~\mbox{mK}$, lower than previously attained in such systems, and this allows us to observe a periodic oscillation in the conductance of the dot as a function of gate voltage. Finite bias and capacitance measurements demonstrate that this oscillation corresponds to a residual quantization of charge, with a renormalized charging energy. We find that the amplitude of the oscillation depends sensitively on both temperature and magnetic field; in particular, the field scale on which the oscillation decreases is that on which time-reversal symmetry is broken. This demonstrates the oscillation is MCB, and reveals how phase coherence leads to the emergence of Coulomb interactions at low temperatures in a system previously thought to be open.
   				
\begin{figure}
\begin{center}
\includegraphics[width=8.0cm, keepaspectratio=true]{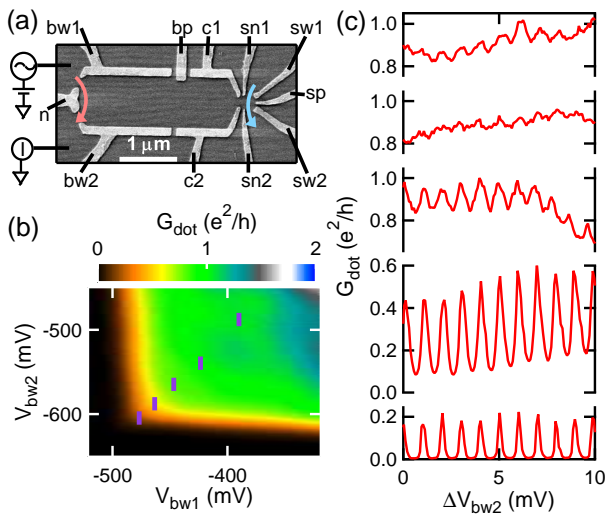}
\end{center}

\caption{(color online) (a) Electron micrograph of a device nominally identical to the measured device. (b) Conductance of the large dot (\AsubB{G}{dot}) at $T= 540~\mbox{mK}$ and $B= 25~\mbox{mT}$. The vertical lines mark the gate voltages at which the cuts in (c) are taken. (c) \AsubB{G}{dot} as a function of $\AsubB{V}{bw2}$ for different settings of the QPCs at $T= 13~\mbox{mK}$ and $B= 25~\mbox{mT}$. The bottom trace is taken at $\AsubB{V}{bw1}= -477~\mbox{mV}$ while the top trace is taken at $\AsubB{V}{bw1}= -390~\mbox{mV}$.  
}
\label{fig:DotPlateau}
\end{figure}	

   We measure a quantum dot fabricated from an epitaxially grown AlGaAs/GaAs heterostructure with a two-dimensional electron gas (2DEG) located at an interface $68$ nm below the surface. The 2DEG has a density of $2 \times 10^{11}~\mbox{cm}^{-2}$ and a mobility of $2 \times 10^6~\mbox{cm}^{2}/\mbox{Vs}$. \Startsubfig{DotPlateau}{a} shows an electron micrograph of the metallic gates we pattern on the surface. Negative voltages are applied to the gates to form a large dot of area $\approx 3~\mu \mbox{m}^{2}$ that we study, as well as an adjacent small dot that is used as a charge sensor \cite{Field1993:NoninvasiveProbe, Schoelkopf1998:RFSET} in the capacitance measurements. The gates bw1, n, and bw2 define the QPCs of the dot, while the gates c1, c2, and bp are used to change the shape of the dot \cite{Chan1995:CondFluc, Rau2010:Dephasing}. Gates c1 and c2 have a small effect on the conductance on of the QPCs, and this effect is compensated by adjusting the gates bw1 and bw2 to maintain the conductance through the QPCs.  For all measurements, the gates sn1 and sn2 are kept sufficiently negative that there is no measurable conductance through the channel between them. We measure the conductance using standard lock-in techniques \cite{footnote:EPAPS}.

\begin{figure}
\begin{center}
\includegraphics[width=8.0cm, keepaspectratio=true]{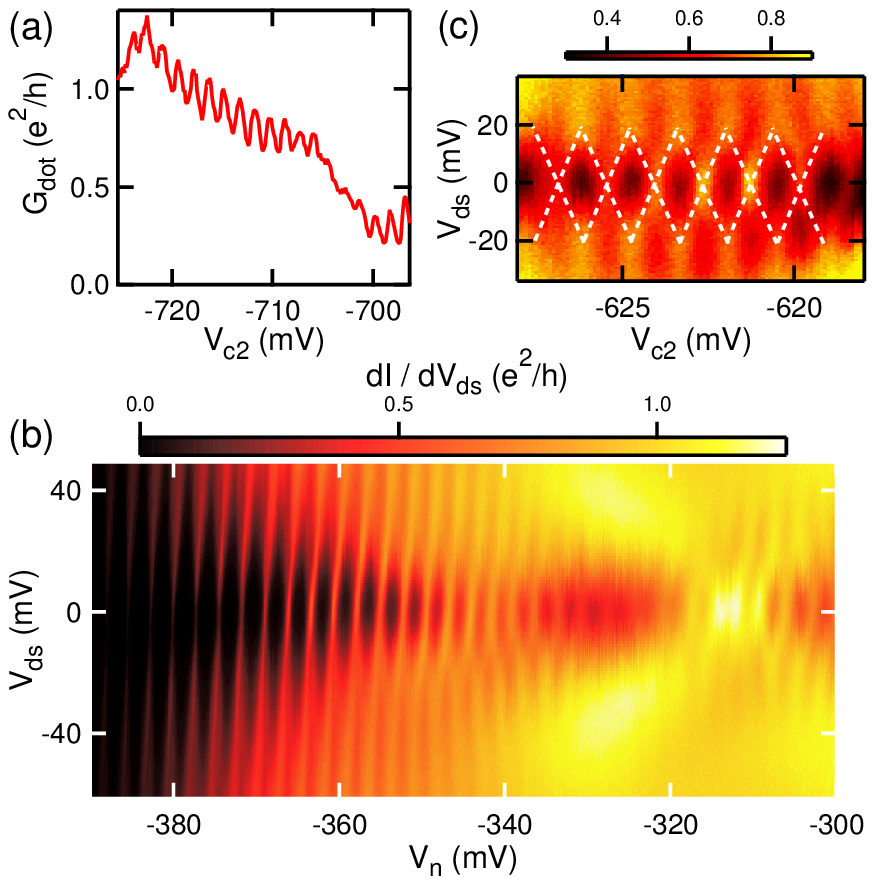}
\end{center}

\caption{(color online) (a) \AsubB{G}{dot} measured at $T= 13~\mbox{mK}$ (charge sensor not active) and $B= 0$ with the gate voltage settings corresponding to the middle of the dot's $1~\esqoh$ plateau. (b) \didv~as a function of \AsubB{V}{ds} and \AsubB{V}{n} at $13~\mbox{mK}$ (with the charge sensor active) and $B= 0$. The QPCs each have a fully transmitting mode at $\AsubB{V}{n} = -315 \mbox{mV}$. (c) \didv~as a function of the voltage on the gate c2, when the conductance through each QPC is kept at 2 \esqoh. The data are taken at $T= 13~\mbox{mK}$ (charge sensor not active) and $B= 0$. The dashed white lines are guides to the eye.   
}
\label{fig:Diamonds}
\end{figure}

   \Startsubfig{DotPlateau}{b} shows the zero-bias conductance of the large dot \AsubB{G}{dot}~as a function of the voltages on the gates bw1 and bw2 (\AsubB{V}{bw1} and \AsubB{V}{bw2}, respectively) that control the two QPCs. These data are taken at  $T= 540~\mbox{mK}$  to suppress  Universal Conductance Fluctuations (UCFs) \cite{Chan1995:CondFluc} and at $B> 5~\mbox{mT}$ to avoid WL, and in this regime \AsubB{G}{dot}~is just the series conductance of the two QPCs. In particular, there is a plateau at $\AsubB{G}{dot}= 1~\esqoh$, corresponding to the 2 \esqoh~plateaus in the conductance of both QPCs \cite{footnote:EPAPS}. \Startsubfig{DotPlateau}{c} shows data taken at $13~\mbox{mK}$ at different values of $(\AsubB{V}{bw1},\AsubB{V}{bw2})$; the gate voltage settings are indicated by the vertical lines in \subfig{DotPlateau}{b}. When the QPCs are in the tunneling regime, clear Coulomb blockade peaks are observed (bottom trace in \subfig{DotPlateau}{c}). However, when both QPCs are set to 2 \esqoh~(top two traces in  in \subfig{DotPlateau}{c}), we observe a residual oscillation in the conductance with the same period as the Coulomb blockade peaks. \Startsubfig{Diamonds}{a} shows this oscillation at $B= 0$ with voltage settings that correspond to the middle of the dot's 1 \esqoh~plateau. The large variations in conductance on the scale of tens of mV in gate voltage are caused by the UCFs, while the rapid periodic oscillation is superimposed on top. The fact that the Coulomb blockade peaks and the oscillation have the same periodicity suggests that they have the same cause: quantization of the charge in the dot.   

    This hypothesis is further supported by measurements of differential conductance \didv~as a function of the bias voltage \AsubB{V}{ds}. \Startsubfig{Diamonds}{b} shows \didv~vs both bias and the voltage on gate n (\AsubB{V}{n}), which controls the conductance of both QPCs. For $\AsubB{V}{n} \lesssim -370~\mbox{mV}$ the QPCs are not fully transmitting and we see clear Coulomb diamonds associated with charge quantization. These diamonds correspond to a charging energy of $110~\mu\mbox{eV}$. As \AsubB{V}{n} is made less negative, the conductance of the QPCs increases. This causes $U$ to be renormalized  \cite{Flensberg1994:DotsConToQPCs, Molenkamp1995:CoulScaling} and as a consequence the vertical size of the diamonds shrinks. At $\AsubB{V}{n} \approx -315 \mbox{mV}$ the QPCs are fully transmitting, and in this regime we see that the oscillations correspond to Coulomb diamond features, with a re-normalized $U^{*} \approx 16 \mu\mbox{eV}$ (see \cite{footnote:EPAPS}). These diamonds are superimposed on larger UCFs which form a Fabry-Perot pattern in gate voltage and bias \cite{Liang2001:FabryPerot}. The diamonds associated with the oscillations are shown in more detail in \subfig{Diamonds}{c}, where the gates are set to the middle of the dot's 1 \esqoh~plateau and the voltage on gate c2 is varied. The presence of Coulomb diamonds support the hypothesis that the oscillations are connected to a residual quantization of charge.

\begin{figure}
\begin{center}
\includegraphics[width=8.0cm, keepaspectratio=true]{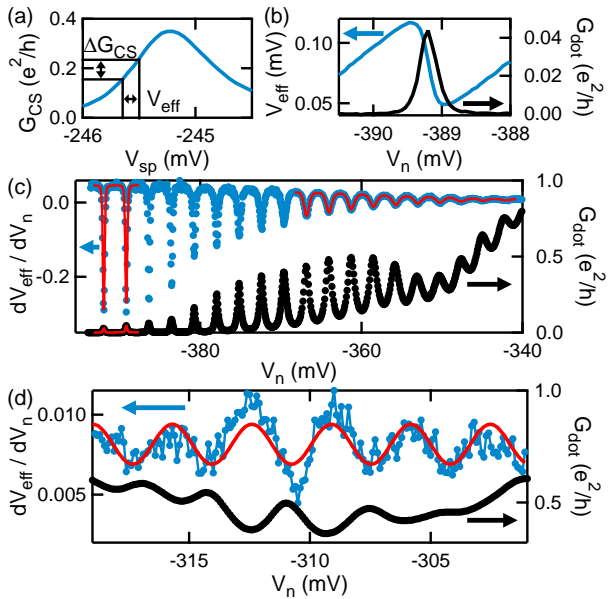}
\end{center}

\caption{(color online) (a) Small dot Coulomb blockade peak used for charge sensing. We  convert the change in the conductance $\Delta G_{CS}$ of the charge sensor into an effective voltage change \AsubB{V}{eff}. (b) Simultaneous measurement of charge sensing signal \AsubB{V}{eff}  (left axis) and transport (right axis) in the large dot, when the conductances of both QPCs are less than 2 \esqoh. (c) Simultaneous measurement of  $d\AsubB{V}{eff}/d\AsubB{V}{n}$ (blue dots, left axis) and transport (black dots, right axis).  The solid line shows fits discussed in the text. (d)  Charge sensing data (left axis) and transport (right axis) at the values of \AsubB{V}{n}~for which the QPCs are open to 2 \esqoh~ conductance. The solid red line is a fit described in the text.
}
\label{fig:Qsen}
\end{figure}  

   To confirm this hypothesis, we directly observe this residual charge quantization with capacitive measurements using the adjacent charge sensor. Making the voltage on a dot gate less negative increases the charge on the dot. The electric fields from both the gate and the additional charge change the conductance of the charge sensor by $\Delta G_{CS}$. We convert $\Delta G_{CS}$ into an effective voltage change \AsubB{V}{eff}, which if applied to the gate sp would produce the same $\Delta G_{CS}$ (\subfig{Qsen}{a}). When the conductances of both QPCs are less than 2\esqoh~we have well-defined Coulomb blockade in  $G_{dot}$ as shown in \subfig{Qsen}{b} (black trace, right axis). A simultaneous measurement of the charge sensor (blue trace, left axis) shows that \AsubB{V}{eff}~initially increases as \AsubB{V}{n} is made less negative because of the capacitance between the gate and the charge sensor. However at the value of \AsubB{V}{n} where an electron is added to the dot, there is a sharp decrease in \AsubB{V}{eff}. To highlight the correspondence between the decrease in \AsubB{V}{eff} and the peaks in \AsubB{G}{dot}, we take the derivative $D = d\AsubB{V}{eff}/d\AsubB{V}{n}$. These data are shown in \subfig{Qsen}{c} over a range in $\AsubB{V}{n}$ that goes from the tunneling to the open regime. For $\AsubB{V}{n}< -375~\mbox{mV}$ the Coulomb blockade peaks are well defined and correspond to large dips in $D$. As \AsubB{V}{n} is increased, the dip size decreases but the dips remain aligned to peaks in \AsubB{G}{dot}. \Startsubfig{Qsen}{d} shows measurements in the range of \AsubB{V}{n} where the QPC conductances are at 2 \esqoh. We see a periodic variation in $D$, with the dips corresponding to peaks in \AsubB{G}{dot}. This measurement confirms that the conductance oscillation corresponds to a residual quantization of charge on the dot.

  We quantitatively analyze these data to estimate the magnitude of the residual quantization. $D$ is determined by the capacitances of the dot $d$ and the charge sensor $CS$ \cite{Berman1999:QuantFlucOfChg}: $D= R_{n} + R_{d}  (C_{n,d} - e~dN_{d}/d\AsubB{V}{n} ) / C_{d,tot}^{*}$. Here $N_{d}$ is the number of electrons on the dot and $R_{n}= C_{n,CS}/C_{sp,CS}$ where $C_{sp,CS}$ and $C_{n,CS}$ are the capacitances of gates sp and n to the charge sensor respectively. Also $R_{d}= C_{d,CS}/C_{sp,CS}$ where $C_{d,CS}$ is the capacitance between the dot and the charge sensor. $C_{n,d}$ is the capacitance of gate n to the dot and $C_{d,tot}^{*}$ is the re-normalized total capacitance of the dot with $U^{*}= e^2/C_{d,tot}^{*}$. For $\AsubB{V}{n}< -385~\mbox{mV}$ the lineshapes are well described by theoretical predictions for $dN_{d}/d\AsubB{V}{n}$ \cite{Schoeller1994:MesoQuantTransport, Grabert1994:ChgFlucSEbox}. The solid red lines in \subfig{Qsen}{c} shows the results of simultaneously fitting the $G_{dot}$ and $D$ data to the theory of Schoeller and Sch\"{o}n, using values of $R_{n}$, $C_{n,d}$, and $C_{d,tot}^{*}$ estimated from other measurements  (see \cite{footnote:EPAPS} for details). This fit gives $R_{d} = 0.93$ (we estimate an error of $\pm 0.21$, see \cite{footnote:EPAPS}), and we use this value to analyze the data in other gate voltage regions. For $-370<\AsubB{V}{n}<-340~\mbox{mV}$ the solid line in \subfig{Qsen}{c} shows a fit to Matveev's prediction for a one-leaded dot without phase coherence, with the adjustable QPC near 2\esqoh~ \cite{Berman1999:QuantFlucOfChg, Matveev1995:CBatPerfectT}. In the limit of a perfectly transmitting contact, this theory predicts there should not be a periodic variation in the charge sensing signal, so we fit the data in \subfig{Qsen}{d} to a model for MCB in a one-leaded dot \cite{Aleiner1998:MCB}: $e~dN_{d}/d\AsubB{V}{n}= C_{n,d} (1+(A/e)\cos(2\pi C_{n,d} \AsubB{V}{n}/e) )$ where $A$ gives the residual charge quantization. We find that $A/e= 0.27^{+0.21}_{-0.08}$, indicating that a significant amount of charge is still quantized. 

   Theoretical results imply that the oscillation depends on phase coherence in the dot. In a two-leaded dot without phase coherence, \AsubB{G}{dot}~should not oscillate when the QPCs are at 2~\esqoh. Even if the QPCs have a small reflection coefficient $r^2$ (defined by $G_{QPC}= 2\esqoh~(1-r^2)$) the lowest order effect is to decrease the average dot conductance, whereas any oscillations are order $r^4$ or higher \cite{Furusaki1995:ThyOfStrongCotun}. The fact that phase coherence is important suggests that the oscillation is MCB. 
   
\begin{figure}
\begin{center}
\includegraphics[width=8.0cm, keepaspectratio=true]{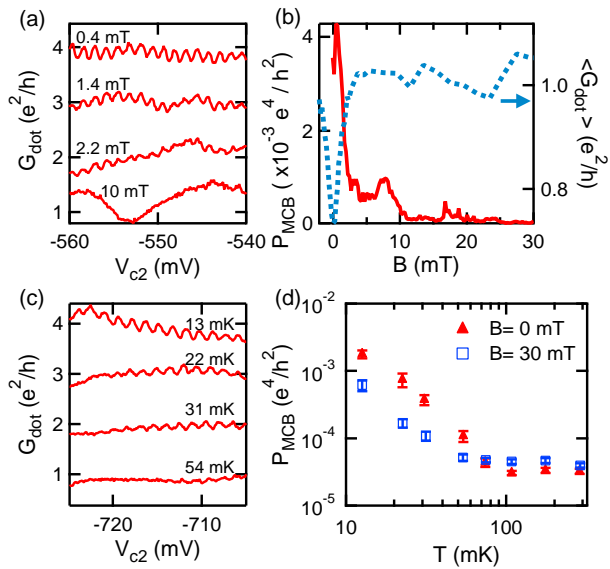}
\end{center}

\caption{(color online) (a) Conductance as a function of \AsubB{V}{c2} at several different magnetic fields at $13 \mbox{mK}$. (b) The solid line (left axis) shows $P_{MCB}$ obtained by Fourier transforming data like those in (a). The dotted line shows the ensemble-averaged conductance of the dot as a function of magnetic field. All data are taken at $13 \mbox{mK}$. (c) Conductance as a function of \AsubB{V}{c2} at $B= 0$ and several different temperatures. (d) $P_{MCB}$ averaged over data taken over a wider range of \AsubB{V}{c2} than in (c) at several different values of \AsubB{V}{c1}.
}
\label{fig:BandTDep}
\end{figure}     

   If the conductance oscillation is MCB, then it should be sensitive to an applied magnetic field which disrupts the constructive interference between time-reversed paths that causes WL. \Startsubfig{BandTDep}{a} shows \AsubB{G}{dot}~as a function of \AsubB{V}{c2}~at several magnetic fields, and the size of the oscillation quickly decreases with increasing field. To quantitatively analyze these data, we follow Cronenwett \etal~\cite{Cronenwett1998:MCBin1ChDots}, Fourier transforming the data and integrating the power spectral density around the frequency of the oscillation to find the power  $P_{MCB}$. The results are shown as the solid line in \subfig{BandTDep}{b}. The dotted line in \subfig{BandTDep}{b} shows \AsubB{G}{dot}~averaged over an ensemble of dot shapes obtained by changing the voltages on gates c1 and c2 \cite{Rau2010:Dephasing, Huibers1998:OpenDotDephasing}. The dip around $B= 0$ is caused by WL, and the width of the dip is the magnetic field scale necessary to break time-reversal symmetry. The fact that the amplitude of the oscillation decreases over the same field scale is strong evidence that the oscillation is MCB. For $B>5~\mbox{mT}$ $P_{MCB}$ is small but non-zero because while the oscillations are weaker and less frequent, they are still present at some gate voltages and magnetic fields, eg. the top two traces in \subfig{DotPlateau}{c}. 

   MCB should also depend sensitively on temperature: the dephasing time, which describes the time scale on which electrons in the dot lose phase coherence, decreases with increasing temperature and interference effects become weaker \cite{Huibers1998:OpenDotDephasing, Rau2010:Dephasing}. \Startsubfig{BandTDep}{c} shows measurements of \AsubB{G}{dot}~at different temperatures, and  \subfig{BandTDep}{d} show the results of extracting $P_{MCB}$ from data at $B= 0$ (filled circles) and $B= 30~\mbox{mT}$ (open squares). The amplitude of the oscillation decreases quickly with increasing temperature (the saturation at $P_{MCB}= 2\times 10^{-5} e^4/h^2$ is from the noise floor). 

In conclusion, we observe an oscillation in the conductance of an open dot that we identify as MCB, a type of Coulomb blockade that depends on electron interference. Previously, a dot with four total modes (one spin degenerate mode in each QPC) was thought to be well described by the theory for the many mode limit, which predicts that MCB should be absent. Our results demonstrate that the understanding of this system, and more generally two terminal mesoscopic systems with several transmitting modes and long coherence times at low temperatures, is incomplete and that theoretical calculations are necessary to explain the interplay of coherence and Coulomb interactions.

	We are grateful to P. W. Brouwer, K. A. Matveev, J. von Delft, Y. Oreg, and I. L. Aleiner for discussions. This work was supported by the NSF under DMR-0906062 and CAREER grant No. DMR-0349354, as well as by the U.S.-Israel BSF grants No. 2008149 and No. 2004278. D.G.-G. thanks the Sloan and Packard Foundations for financial support, and acknowledges a Research Corporation Research Innovation grant.

\newpage

\renewcommand{\thefigure}{S\arabic{figure}}
\renewcommand{\theequation}{S\arabic{equation}}
\setcounter{figure}{0}

\noindent \begin{center}{\Large Supplementary Information for Coulomb Blockade in an Open Quantum Dot}\par\end{center}{\Large
\par}

\section{Quantum Dot and Charge Sensor Conductance Measurements}
	
	We measure the large quantum dot by placing a small oscillating voltage on top of the dc bias voltage and measuring the resulting current with a DL Instruments Model 1211 current pre-amplifier and a Princeton Applied Research 124A lock-in amplifier (the circuit for the large dot is sketched in Fig. 1(a) in the main text). For the measurements of the large dot we use an oscillation frequency of $17$ Hz and an excitation voltage $V_{exc}$ from $1$ to $5~ \mu\mbox{Vrms}$. We use a separate but identically constructed circuit to measure the charge sensor. For the charge sensor measurements we use a frequency of $97$ Hz and an excitation voltage of $5~\mu\mbox{Vrms}$. 
	
\section{QPC conductance plateaus}

\begin{figure}[h!]
\begin{center}
\includegraphics[width=8.0cm, keepaspectratio=true]{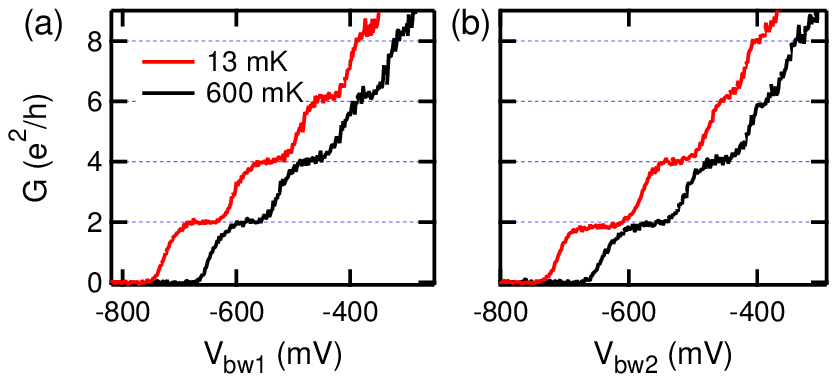}
\end{center}

\caption{(a) Conductance measurements of QPC 1 (a) and QPC 2 (b) at $13~\mbox{mK}$ and $600~\mbox{mK}$. The $600~\mbox{mK}$ plateaus have been shifted horizontally for clarity. 
}
\label{fig:Plateaus}
\end{figure}	  

   Measurements of the individual QPCs show clear conductance plateaus quantized at integer multiples of $2~\esqoh$, demonstrating that the QPCs do not have spurious resonances that could cause the observed oscillation in the dot conductance. \Startsubfig{Plateaus}{a} shows the conductance of QPC 1, formed by the gates bw1 and n, with no voltage applied to any other dot gates. These data show plateaus at $G= 2~\esqoh$ and $4~\esqoh$. \Startsubfig{Plateaus}{b} shows similar data for QPC 2, with voltages applied only to the gates bw2 and n. The clear plateaus in these data show there is no evidence of spurious resonances. The figure also shows the results of measuring the QPCs at $600~\mbox{mK}$. The slope of the increase in conductance between the plateaus at low and high temperatures is approximately equal, indicating that the dominant energy scale for the opening of a new mode in the QPC is greater than $\approx 50~\mu\mbox{eV}$.
      
   For the measurements in the main text, the device was cooled to $4~\mbox{K}$ with a positive bias of $\approx +200~\mbox{mV}$ applied to all gates. The positive bias ``pre-depletes'' the gates, preventing us from characterizing the individual QPCs even if all other gates are set to $0~\mbox{V}$. The measurements in \subfig{Plateaus}{a} and (b) have been performed after all the dot measurements reported in the paper, and following a partial thermal cycle to a temperature on the order of or greater than $100~\mbox{K}$ to reduce the effects of the positive bias voltage applied when the dot was initially cooled down to $4~\mbox{K}$. The measurements in \subfig{Plateaus}{a} and (b) were taken with a small excitation voltage and little averaging, which cause the noise on the measurement.

\section{Estimation of the reflection coefficient}
   
   Through careful analysis of the temperature dependence of the average dot conductance at finite magnetic field we have determined that the reflection coefficients of the QPCs are on the order of $2\%$ or less. The analysis is discussed below.
   
   For the work reported in this paper we have carefully tuned both QPCs to their 2~\esqoh~plateaus so that there is one fully transmitting spin-degenerate mode in each QPC. However, even at the optimal QPC settings there may still be a small reflection coefficient in the QPCs. This reflection coefficient affects the dot conductance, and this can be observed in the conductance at finite magnetic field where weak localization is absent. To extract the conductance at finite field, we analyze measurements of the ensemble averaged dot conductance as a function of magnetic field like that shown by the dotted line in Fig. 4(b) of the main text. In this data we see the weak localization dip at B= 0, and following Huibers \etal \cite{Huibers1998:OpenDotDephasing} we fit this dip to a Lorentzian:
   
\begin{displaymath}   
	<G_{dot}(B)> = <G_{dot}>_{B\neq 0} - \frac{A}{1+(2B/B_{c})^2}
\end{displaymath}
where $<G_{dot}>_{B\neq 0}$ is the average conductance at finite field, and $A$ and $B_{c}$ are the size and width of the weak localization dip, respectively. The temperature dependence of $<G>_{B\neq 0}$ depends on the reflection coefficient of the QPCs. 

\begin{figure}
\begin{center}
\includegraphics[width=7.8cm, keepaspectratio=true]{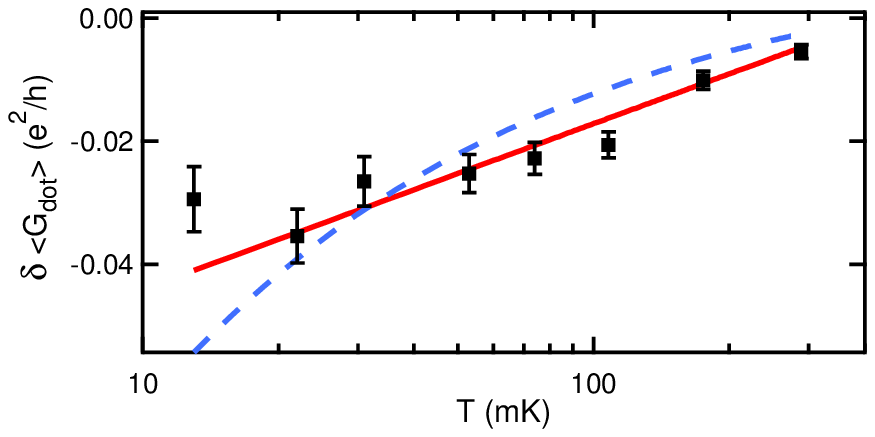}
\end{center}

\caption{Temperature dependence of the change in the ensemble averaged dot conductance at finite magnetic field. The solid and dashed lines are fits discussed in the text. 
}
\label{fig:rQPC}
\end{figure}	  

   To characterize the temperature dependence, we find the difference 
\begin{displaymath}
	 \delta <G_{dot}> = <G_{dot}>_{B\neq 0}(T) - <G_{dot}>_{B\neq 0}(T_{0})
\end{displaymath}
where $T_{0}= 435~\mbox{mK}$ and this quantity is plotted in \allfig{rQPC}. We see there is a very small temperature dependence of the dot conductance. There is no explicit theoretical prediction for the temperature dependence of a coherent quantum dot with one fully transmitting spin-degenerate mode in each QPC (this is the $N=4$ case, where $N$ is the total number of transmitting channels in both QPCs). However, for a coherent dot with $N\gg1$ Brouwer \etal~\cite{Brouwer2005:NonequilCBThy} have calculated the temperature dependence and find that it is the same as that for an incoherent dot\cite{Golubev2004:ElecTransportThruQD} with $N \gg 1$ and is given by:
\begin{equation}
	\delta <G_{dot}> = \frac{e^2}{h} \left[ \frac{-r^2}{2} \ln{\frac{T_{0}}{T}} \right] 
\label{eq:dGGZ}
\end{equation}
In this equation the reflection coefficients $r^2$ of the two QPCs are assumed to be equal and are defined by $G_{QPC}= 2~\esqoh~(1-r^{2})$. The solid red line in \allfig{rQPC} shows the results of fitting the data to this equation, and it appears the $N\gg 1$ theoretical prediction gives a decent fit for $N= 4$. From the fit we obtain $r^2 \approx 0.02$. 

   Furusaki and Matveev \cite{Furusaki1995:ThyOfStrongCotun} have calculated the temperature dependence for an incoherent dot with $N=4$ and found
\begin{equation}
	\delta <G_{dot}> = \frac{e^2}{h} \left[ \frac{-4 r^2 \Gamma(\frac{3}{4})}{\Gamma(\frac{1}{4})} \sqrt{\frac{\gamma U}{k_{B}\pi}} \right] (T^{-1/2} - T_{0}^{-1/2})
\label{eq:dGMat}
\end{equation}
In this equation $\gamma= \exp(C)$ where $C= 0.5772\ldots$. The dashed blue line shows a fit to this equation and we obtain $r^2 \approx 0.007$. We note that the theoretical prediction for a coherent dot fits the data better than the prediction for the incoherent dot, indicating the importance of phase coherence to understanding the dot properties. These fits allow us to conclude that the reflection coefficients of the QPCs are small (on the order of $2\%$ or less) and hence effects that are higher order in $r^2$ should be suppressed.

\section{Determining the renormalized charging energy}

  In this section, we describe how we analyze the data in figure 2(b) in the main text to extract the renormalized charging energy $U^{*}\approx 16~\mu\mbox{eV}$ near $\AsubB{V}{n}= -315~\mbox{mV}$. This value is used to determine $C_{d,tot}^{*}= e^2/U^{*}$ which is an input for the fit in figure 3(d) of the main text.

\begin{figure}[h!]
\begin{center}
\includegraphics[width=8.0cm, keepaspectratio=true]{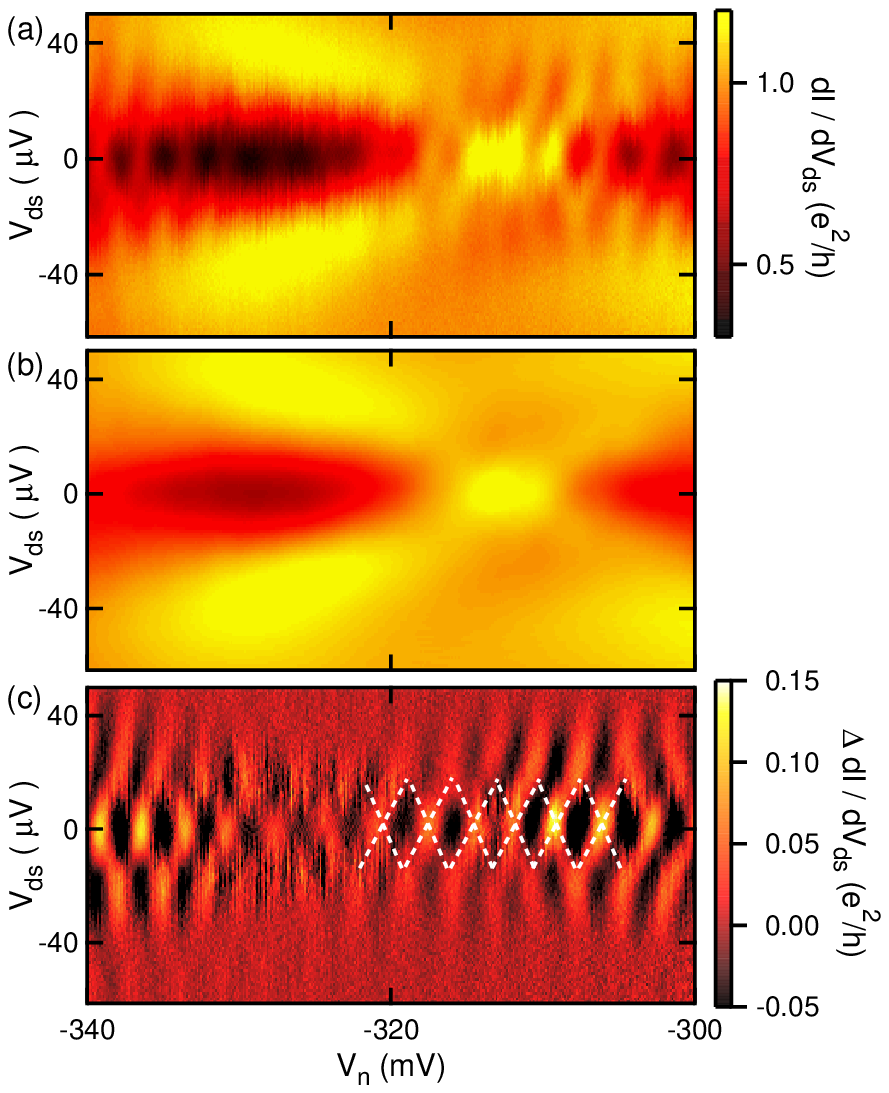}
\end{center}

\caption{(a) Conductance as a function of \AsubB{V}{n} at $13~\mbox{mK}$ (with charge sensor active) and $B= 0$. The QPCs each have a fully transmitting mode at $\AsubB{V}{n}= -315~\mbox{mV}$. (b) Result of averaging (a) over $\Delta \AsubB{V}{n}\approx 3~\mbox{mV}$, which is one period of the Coulomb diamonds. The averaging leaves only the background Fabry-Perot resonance. (c) Result of subtracting the background in (b) from the data in (a). These subtracted data allow us to identify Coulomb diamonds (dashed white lines are guides to the eye) and to extract a renormalized charging energy $U^{*}$.
}
\label{fig:renormU}
\end{figure}

  In figure 2(b) of the main text the diamonds are on top of a background conductance caused by Fabry-Perot interference of electrons in the big dot, making determination of the charging energy more difficult. The procedure for subtracting this background is demonstrated in \allfig{renormU}.

  \Startsubfig{renormU}{a} shows the data from figure 2(b) in the main text. To isolate the background Fabry-Perot pattern, we smooth the data by averaging in gate voltage over the period of the Coulomb diamonds, $\Delta \AsubB{V}{n}\approx 3~\mbox{mV}$. The result of this averaging is shown in \subfig{renormU}{b}. We then subtract these averaged data from the raw data in \subfig{renormU}{a} to isolate the oscillation. The result is shown in \subfig{renormU}{c}, with dashed white lines as guides to the eye. From the Coulomb diamonds, we find a renormalized charging energy $U^{*} \approx 16~\mu\mbox{eV}$ at $\AsubB{V}{n}= -315~\mbox{mV}$.

\section{Charge Sensing Fits}
\label{sec:Fits}

   In this section we describe how we fit the charge sensing data in the range $\AsubB{V}{n}< -385~\mbox{mV}$ in figure 3 of the main text to determine the capacitance ratio $R_{d} \approx 0.93 \pm 0.21$. This ratio is used as an input for the fits in the other two \AsubB{V}{n}~ranges discussed in the main text.

\begin{figure}[h!]
\begin{center}
\includegraphics[width=8.0cm, keepaspectratio=true]{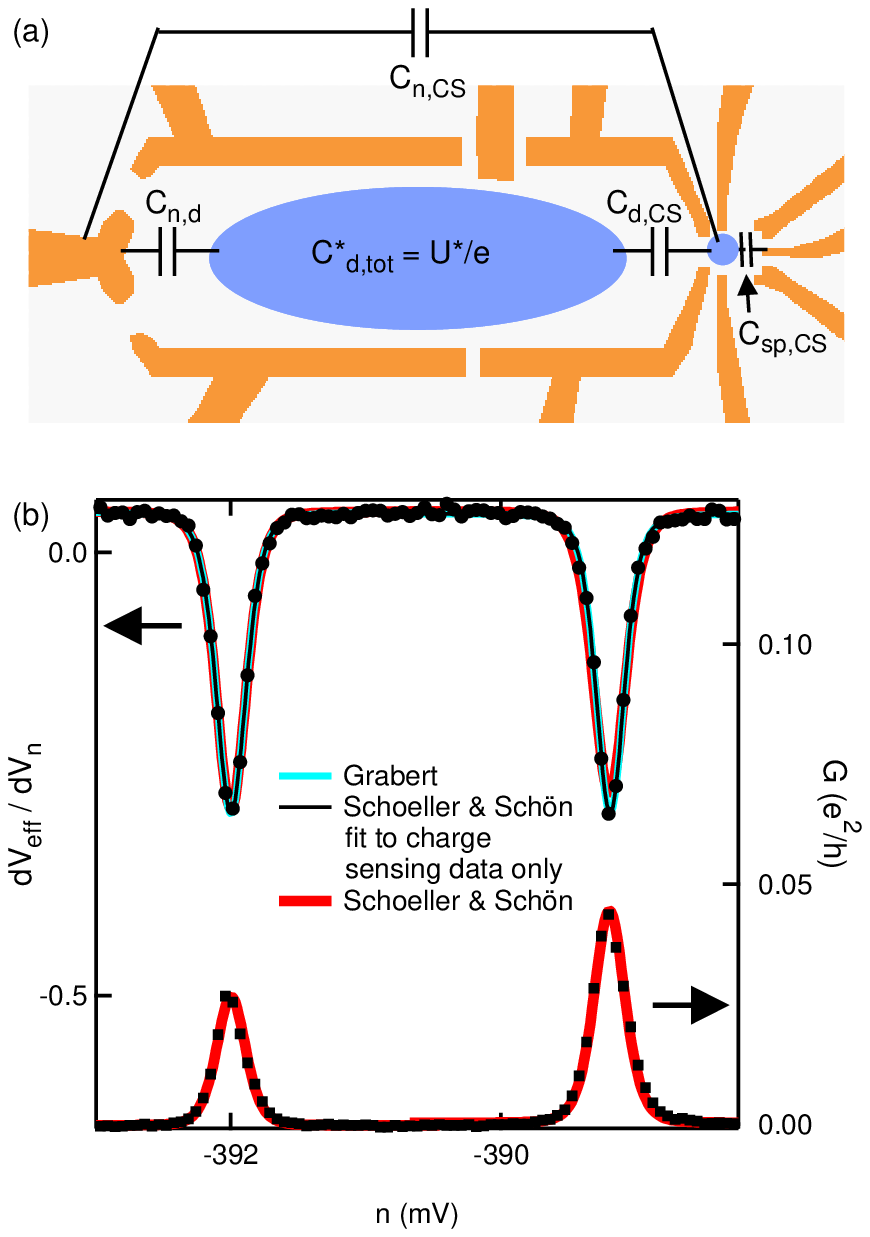}
\end{center}

\caption{(a) Schematic of the large quantum dot and charge sensor, showing the capacitances relevant to determining the charge sensing signal as described by equation 1. (b) Simultaneous measurement of transport and charge sensing. The solid lines are fits discussed in the text.
}
\label{fig:ChgSenFits}
\end{figure}

   To fit the charge sensing data in figure 3 in the main text we use a model similar to that in Berman \etal~\cite{Berman1999:QuantFlucOfChg} illustrated in \subfig{ChgSenFits}{a}. In this diagram, $C_{sp,CS}$ and $C_{n,CS}$ are the capacitances of gates sp and n to the charge sensor, respectively. $C_{n,d}$ is the capacitance of gate n to the large dot, $C_{d,CS}$ is the capacitance between the dot and the charge sensor, and $C_{d,tot}^{*}$ is the re-normalized total capacitance of the large dot, related to the re-normalized charging energy by $U^{*}= e^2/C_{d,tot}^{*}$. Based on this model for two quantum dots \cite{vanderWiel2002:DQDreview} (the large dot and the charge sensor), we can derive the dependence of $D= d\AsubB{V}{eff}/d\AsubB{V}{n}$ on the capacitances, giving rise to the equation 

\begin{equation}   
   D= R_{n} + R_{d} \frac{C_{n,d} - e~dN_{d}/d\AsubB{V}{n}}{C_{d,tot}^{*}}
\label{eq:Deqn}
\end{equation}   

given in the main text, with $R_{n}= C_{n,CS}/C_{sp,CS}$ and $R_{d}= C_{d,CS}/C_{sp,CS}$.

   \Startsubfig{ChgSenFits}{b} shows a simultaneous measurement of the charge sensing signal and conductance for $\AsubB{V}{n}< -385~\mbox{mV}$ (magnification of the data in figure 3c of the main text). In this region, a calculation of the individual QPC conductances based on measurements of the dot conductance show that one of the dot QPCs is partially transmitting, while the other is in the tunneling regime. For this configuration, theoretical work by Schoeller and Sch\"{o}n \cite{Schoeller1994:MesoQuantTransport} and Grabert \cite{Grabert1994:ChgFlucSEbox} predict $dN_{d}/d\AsubB{V}{n}$. We fit the data in \subfig{ChgSenFits}{b} to these theoretical predictions using equation \ref{eq:Deqn}. For these fits, we input $C_{n,d}\approx 58~\mbox{aF}$ estimated from the spacing of the Coulomb Blockade peaks and $U^{*}\approx 115~\mu eV$ estimated from the Coulomb diamonds in figure 2(a) of the main text. For the Schoeller and Sch\"{o}n fits we input $g=G_{QPC1}+G_{QPC2}\approx 0.8~\esqoh$ estimated from the calculation of the QPC conductances (the fits do not depend sensitively on this estimate). To estimate $R_{n}$, we use the fact that in the region $-330~\mbox{mV} < \AsubB{V}{n} < -300~\mbox{mV}$ when both QPCs are open, $d N_{d}/d\AsubB{V}{n}$ is close to its classical value $C_{n,d}/e$, with small perturbations from the residual charge quantization. So the average charge sensing signal in this region is $<D> \approx R_{n}$, and from the data we estimate $R_{n} \approx 0.008$. This value is close to what we expect: from measurements of the charge sensor response to voltage changes on individual gates, we measure $C_{n,CS}= 0.08~\mbox{aF}$ and $C_{sp,CS}= 12~\mbox{aF}$, giving $R_{n}= 0.007$. 
   
   Using these values as inputs we fit the data in \subfig{ChgSenFits}{b}. The solid red line shows a simultaneous fit to both the charge sensing data and the transport data from the theory of Schoeller and Sch\"{o}n, a fit also shown in figure 3(c) of the main text. The solid black line shows a fit of the same theory to only the charge sensing data. Finally, the solid blue line shows a fit to the theoretical prediction of Grabert. The theoretical predictions agree well with the data, and from these and other fits, we extract $R_{d} \approx 0.93 \pm 0.21$. A value of $R_{d}$ near $1$ is reasonable, because the large dot essentially ``gates'' the charge sensor, and so we expect it to have a capacitance comparable to one of the charge sensor's gates. Using $C_{sp,CS}= 12~\mbox{aF}$ estimated from measurements of the charge sensor, we obtain $C_{d,CS}\approx 11~\mbox{aF}$.
   
   The fits return temperatures $T\approx 54 \pm 20~\mbox{mK}$, which are significantly higher than our electron temperature of $13~\mbox{mK}$. The high temperatures extracted from the fits are caused by the Coulomb blockade peaks being broadened by the back-action of the charge sensor on the large quantum dot\cite{Turek2005:SETbackaction}. When an electron tunnels on and off the charge sensor, it ``gates" the large dot and acts like a fluctuating gate voltage that affects the energy of the large dot. The fluctuations of the large dot energy have energy scale  $(C_{d,CS}/C_{d,tot}^{*}) (e^2/C_{s,tot})= (C_{d,CS}/C_{s,tot}) U^{*}\approx 8~\mu\mbox{eV}$, where $U^{*}= 115~\mu\mbox{eV}$ in this gate voltage range, and $C_{s,tot}\approx 150~\mbox{aF}$ is the total capacitance of the small dot. These fluctuations broaden the Coulomb Blockade peaks, and this broadening appears as an increased temperature in our fits: the energy scale of $8~\mu\mbox{eV}$ converts (via Boltzmann's constant) to a temperature of $93~\mbox{mK}$, which is on the order of the broadening that we obtain in the fits. We note that as we increase the conductance of both QPCs in the large dot, the re-normalized charging energy $U^{*}$ decreases and the effect of the back-action on the large dot grows smaller. When both QPCs are fully transmitting, $U^{*}\approx 16~\mu\mbox{eV}$ and the energy scale of the back-action is $\approx 1~\mu\mbox{eV}$, which is on the order of our temperature of $13~\mbox{mK}$.


\newcommand{\noopsort}[1]{} \newcommand{\printfirst}[2]{#1}
  \newcommand{\singleletter}[1]{#1} \newcommand{\switchargs}[2]{#2#1}

\end{document}